# Transport and magnetic $J_c$ of $MgB_2$ strands and small helical coils

M.D. Sumption[1], M. Bhatia[1], M. Rindfleisch[2],

M. Tomsic[2], and E.W. Collings[1]

[1] LASM, Materials Science and Engineering Department,

OSU, Columbus, OH 43210, USA

[2] Hyper Tech Research, Inc. Columbus, OH 43210, USA

**Abstract**

The critical current densities of $MgB_2$ monofilamentary strands with and without SiC additions were measured at 4.2 K. Additionally, magnetic $J_c$ at $B = 1$ T was measured from 4.2 K to 40 K. Various heat treatment times and temperatures were investigated for both short samples and small helical coils. SiC additions were seen to improve high field transport $J_c$ at 4.2 K, but improvements were not evident at 1 T at any temperature. Transport results were relatively insensitive to heat treatment times and temperatures for both short samples and coils in the 700°C to 900°C range.

**Introduction**

Many groups now fabricate $MgB_2$ wires by the powder-in-tube (PIT) process, typically using Fe, Cu, Cu-Ni, monel, or stainless steel (SS) as the outer sheath. There are two main variants of PIT $MgB_2$ fabrication, ex-situ and in-situ, each of which has advantages and disadvantages.

Grasso[1,2] filled Ni tubes with ex-situ $MgB_2$ powders and cold worked them by groove rolling, drawing, and rolling. Heat treatment (HT) conditions experienced by these in-situ route tapes were 800-940°C for several hours. These HTs were intended mainly to recover $T_c$ depressed during the deformation, and to sinter the powders. Measurable $J_c$s were obtained out to 12 T in some cases (about 400 A/cm$^2$ 12 T was the best high field result of[1]) and the irreversibility field, $H_{irr}$, was slightly higher than 12 T for fields applied perpendicular to the tape face. An anisotropy factor of about 1.4 in $H_{irr}$ has been reported for tapes, with $H_{irr,//} > H_{irr,\perp}$ indicating some degree of crystalline alignment[2]. Goldacker et al[3] have investigated ex-situ route powders in SS tubes with Fe and Nb barriers. Flükiger, Suo, et al[4], have fabricated seven-stack multifilaments using ex-situ powders, giving attention to powder sizes[4]. Ball milling of $MgB_2$, a method extensively employed by Flükiger[5,6], was seen to increase $H_{irr}$ and $J_c$. Matsumoto and Kumakura[7,8] investigating various hardness sheaths for ex-situ powder tapes, confirmed that hard outer sheaths, e.g., SS, were required to encourage proper sintering of the powders and good $J_c$ values.

Moving to the in-situ method, Fujii and Kumakura[9] by introducing tungsten boride (WB) into in-situ-powder-route Fe-sheathed $MgB_2$ tapes, decreased the field dependence of $J_c$ as compared to their undoped samples. Their interpretation in terms of WB-induced pinning enhancement was supported by $F_p$ curve analysis. Matsumoto and Kumakura[10] by doping in-situ process $MgB_2$ with $SiO_2$ and SiC improved both $J_c$ (to between 1600 and 6500 A/cm$^2$ at 12 T) and $H_{irr}$. Matsumoto and Kumakura[11] also introduced $ZrSi_2$, $ZrB_2$, and $WSi_2$ additions to good effect. In a series of papers, Dou et al [12,13,14] have studied the effects of nano-Si doping[14], nano-carbon doping[15], and carbon nanotube doping[16,17] on the properties of $MgB_2$ strands. Glowacki[18,19] has also been pursuing PIT $MgB_2$, with useful demonstrations particularly with regard to cables and shielding.

Below we report on the properties of a series of wires, mostly of SiC-doped $MgB_2$. The samples measured were taken from longer batch runs, part of a scale-up effort under way at Hyper Tech Research Inc. Transport results are presented for strands with and without SiC at 4.2 K at high fields, and improvement is seen in response to the SiC additions. Magnetic $J_c$ results at higher temperatures show less difference with doping. The effect of various HT protocols is then investigated, and the reaction window is shown to be relatively large for the present wires. Small helical coils HTd at various temperatures consistently yielded transport $J_c$s as high as $10^4$ A/cm$^2$ at 8 T.

**Experimental**

*Sample preparation*

The 99.9% pure starting Mg powders were -325 mesh, and the 99.9% pure B powders were amorphous, with a typical size of 1–2 µm. The powders were V-mixed and then run in a planetary mill. Two powder types were made: pure binary powders and those with SiC additions. The SiC was added during the V-mixing stage, in the ratio of 10 mole % SiC to 90 mole % of binary $MgB_2$; that is, $[(MgB_2)_{0.9}(SiC)_{0.1}]$. The SiC had an average particle diameter of about 0.2~0.3 µm, with maximum particle size of nearly 0.5 µm. Table 1 lists the strands (monofilamentary) and whether or not SiC was present as well as other details.

The PIT strands were processed at Hyper Tech Research (HTR) by the so-called continuous tube filling/forming (CTFF) process in which the powder is dispensed onto a strip of metal as it is being continuously formed into a tube at about 1 m/min. For $MgB_2$ strands the strip is high purity Fe (23 mm wide, 0.13 mm thick). The result of CTFF processing is an overlap-closed, powder-filled tube or "sheath". After exiting the mill at a diameter of 5.9 mm the filled, overlap-closed tube is inserted into a seamless monel or Cu-30Ni tube and wire drawn to final size. For further details of this process see[20].

Heat treatments were then performed under Ar. Ramp-up times were typically 30 min, and the samples were furnace cooled. The times and temperatures at the plateau were 5-30 minutes at 675-850°C.

*Measurements*

Four-point transport $J_c$ measurements were made on two sample types, short samples and small helical coils (ITER barrel samples[21]). The short samples were 3 cm in length, with a gauge length of 5 mm). The coils held a single layer of wire, with each turn separated from its neighbors. The ends of the 1 m coil were soldered onto Cu end-rings which formed the current contacts. The voltage taps were 50 cm apart. Standard Pb–Sn solder was used for forming the contacts, and the $J_c$ criterion was 1 µV/cm. Most measurements were made at 4.2 K in background fields of up to 15 T (applied transversely to the strand). In a few cases, vibrating sample magnetization (VSM) measurements were made on short segments of strand. The VSM system had a 1.7 T maximum field, and loop measurement times were typically 2-10 min in duration. Measurements were performed from 4.2 K to 40 K. Fields were applied transverse to the strand, which was 1 cm in length. Magnetic $J_c$ results were extracted using the Bean critical state model relevant for cylindrical samples.

**Results**

Figure 1 which displays $J_c$ vs $B$ for various short lengths of strand shows that the 4.2 K high field performance is increased by the SiC additions, see also[22]. However, Figure 2 shows no real difference in the magnetic $J_c$s at lower fields (1T). This might be expected since the primary role of SiC is to increase the $H_{irr}$ and $H_{c2}$. Nevertheless some increases in pinning have also been seen[14-17], but not in the present case at higher temperatures and lower fields..

Figure 3 displays the $J_c$ vs $B$ curves at 4.2 K for short lengths of SiC containing strand HTd at various schedules. All but the least aggressive HT (700°C/5 min) gave comparable results, indicating that all reactions are complete beyond that point. At 8 T, these strands have $J_c$s of about $10^4$ A/cm$^2$. Figure 4 shows the transport $J_c$ at 4.2 K of similar strands wound into small helical coils. These samples also give $10^4$ A/cm$^2$ at 8 T. Since these strands have not yet been provided with a stabilizing sheath they are unstable at fields below this point.

**Conclusions**

In-situ, SiC doped MgB$_2$ strands in Fe/CuNi sheaths have been tested for transport $J_c$ at 4.2 K, and compared to similar strands without SiC. SiC was shown to improve the high field transport properties. Magnetic $J_c$ measurements, taken at 1 T and at temperatures between 4 and 40 K did not show any significant improvement. Short samples of SiC doped strands and small helical coils HTd for various times at 700, 800, and 900°C gave similar results provided the HT duration was longer than 5 minutes. Transport $J_c$ at 8 T and 4.2 K was seen to be $10^4$ A/cm$^2$ for both short samples and small coil samples.

**Acknowledgements**

This work was supported by a State of Ohio Technology Action Fund Grant, by the US Department of Energy, Division of High Energy Physics, Grant No. DE-FGG02-95ER40900, and by the US Air Force, Grant No. F33615-03-C-2344.

6.  R Flükiger, H.L. Suo, N. Musolino, et al., Physica C **385**, 286–305 (2003).

7.  H. Kumakura, A. Matsumoto, H. Fujii, H. Kitaguchi, K. Togano, Physica C **382,** 93–97 (2002).

8.  A. Matsumoto, H. Kumakura, H. Kitaguchi, H. Fujii, K. Togano, Physica C **382**, 207–212 (2002).

9.  H Fujii, K Togano and H Kumakura,", Supercond. Sci. Technol. **16**, 432–436 (2003).

10. A. Matsumoto, H. Kumakura, H. Kitaguchi, and H. Hatakeyama, Supercond. Sci. Technol. **16**, 926–930 (2003).

11. Y. Ma, H. Kumakura, A. Matsumoto, et al., Supercond. Sci. Technol. **16**, 852–856 (2003).

12. S. Soltanian, X.L. Wang, I. Kusevic, et al., Physica C **361**, 84-90 (2001).

13. S. Soltanian, X.L. Wang, A.H. Li, et al., Solid State Commun. **124,** 59-62 (2002).

14. S.X. Dou, A.V. Pan, S. Zhou, et al., Supercond. Sci. Technol. **15**, 1587–1591 (2002).

15. A.V. Pan, S. Zhou, H. Liu and S. Dou, Supercond. Sci. Technol. **16,** 639–644 (2003).

16. S.X. Dou, A.V. Pan, S. Zhou, et al., Journal of Appl. Phys. **94**, 1850-1856 (2003).

17. S.X. Dou, W.K. Yeoh, J. Horvat, et al., Appl. Phys. Lett. **83**, 4996-4998 (2003).

18. B A Glowacki, M Majoros, M Vickers, et al., Supercond. Sci. Technol. **16,** 297–305 (2003).

19. B.A. Glowacki, M. Majoros, M. Eisterer, et al., Physica C **387**, 153–161 (2003).

20. V. Selvamanickam et al., Supercond. Sci. Technol. **8,** 587-590 (1995).

**List of Tables**

Table. 1. Strand Specifications.

**List of Figures**

Figure 1. Transport $J_c$ vs $B$ at 4.2 K for short samples with and without SiC doping and given various HTs.

Figure 2. Magnetic $J_c$ (1T) vs T for short samples with and without SiC doping and given various HTs.

Figure 3. Transport $J_c$ vs $B$ at 4.2 K for short samples with SiC doping given various HTs.

Figure 4. Transport $J_c$ vs $B$ at 4.2 K for small helical coil samples with SiC doping given various HTs.

Table 1. Strand Specifications.

| Name | Tracer ID | SiC% | HT (°C/min) | SC% | OD, mm |
|---|---|---|---|---|---|
| **Set I – SiC/No SiC, Cu-15Ni** | | | | | |
| SiC700/30 | 256a | 10 | 700/30 | 27.5 | 0.827 |
| SiC800/15 | 256a1 | 10 | 800/15 | 27.5 | 0.827 |
| SiC800/05 | 256a2 | 10 | 800/05 | 27.5 | 0.827 |
| NSC700/30 | 253b | 0 | 700/30 | 24.1 | 0.791 |
| NSC800/15 | 253b1 | 0 | 800/15 | 24.1 | 0.791 |
| NSC700/15 | 253b2 | 0 | 700/15 | 24.1 | 0.791 |
| **Set II – SiC time and temp, Cu-30Ni** | | | | | |
| SS700/05 | 262 | 10 | 700/05 | 27 | 0.83 |
| SS700/20 | 262 | 10 | 700/20 | 27 | 0.83 |
| SS800/05 | 262 | 10 | 800/05 | 27 | 0.83 |
| SS800/20 | 262 | 10 | 800/20 | 27 | 0.83 |
| SS900/05 | 262 | 10 | 900/05 | 27 | 0.83 |
| **Set III – Small helical coils, Cu-30Ni** | | | | | |
| HCOIL700/30 | 362 | 10 | 700/30 | 28 | 0.83 |
| HCOIL800/10 | 362 | 10 | 800/30 | 28 | 0.83 |
| HCOIL900/30 | 362 | 10 | 900/30 | 28 | 0.83 |

[a] 10 mol% of SiC added to 90 mol% of binary MgB2 [(MgB2)0.9(SiC)0.1]. SiC used was ≈ 200 nm.

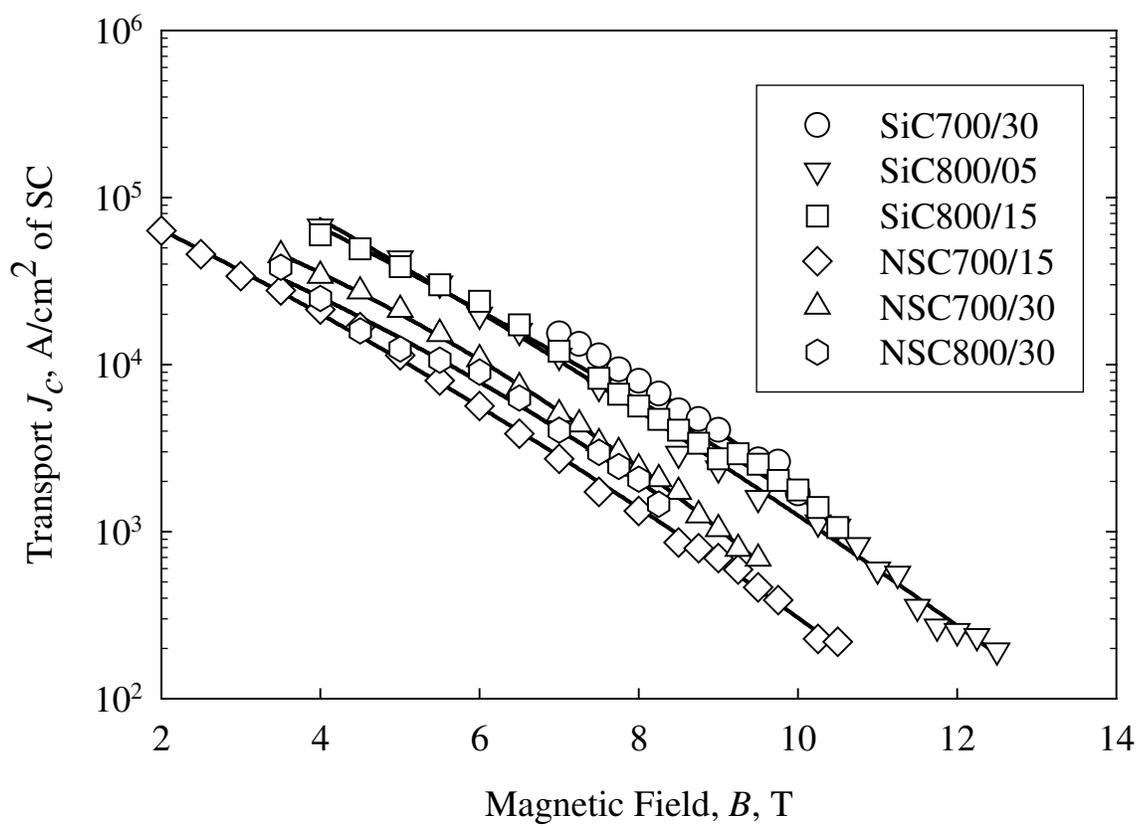

**Figure 1 SUMPTION**

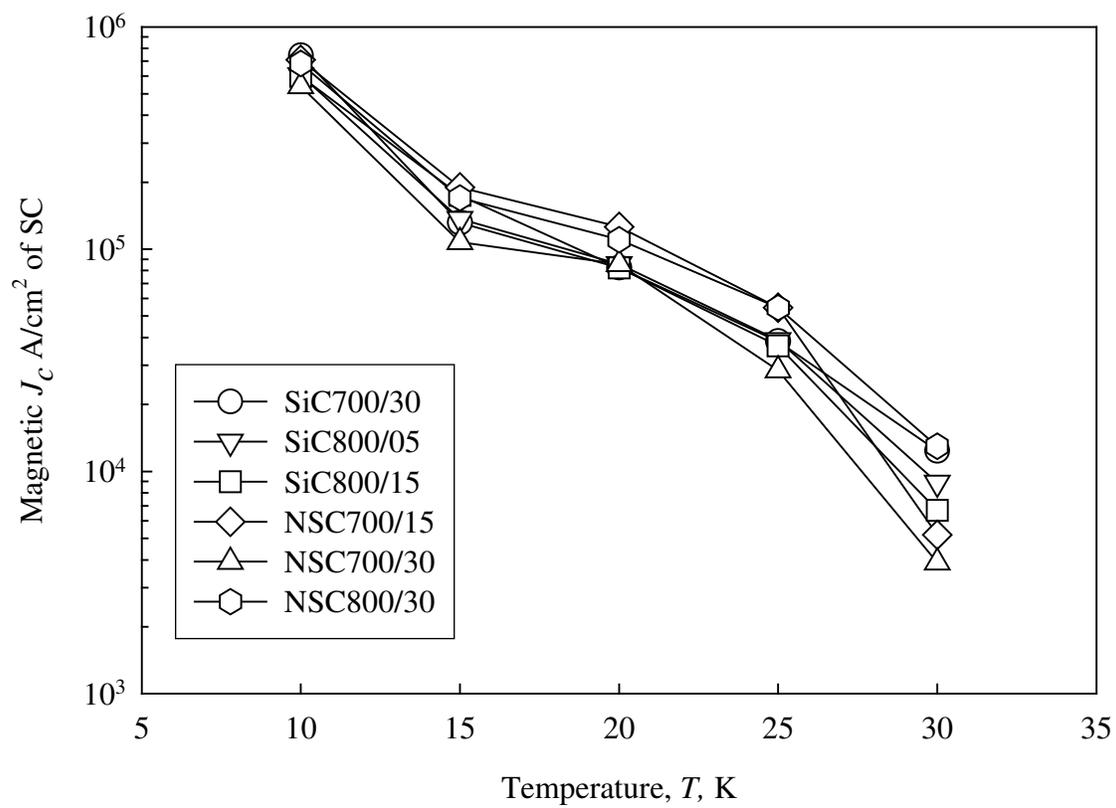

**Figure 2. SUMPTION**

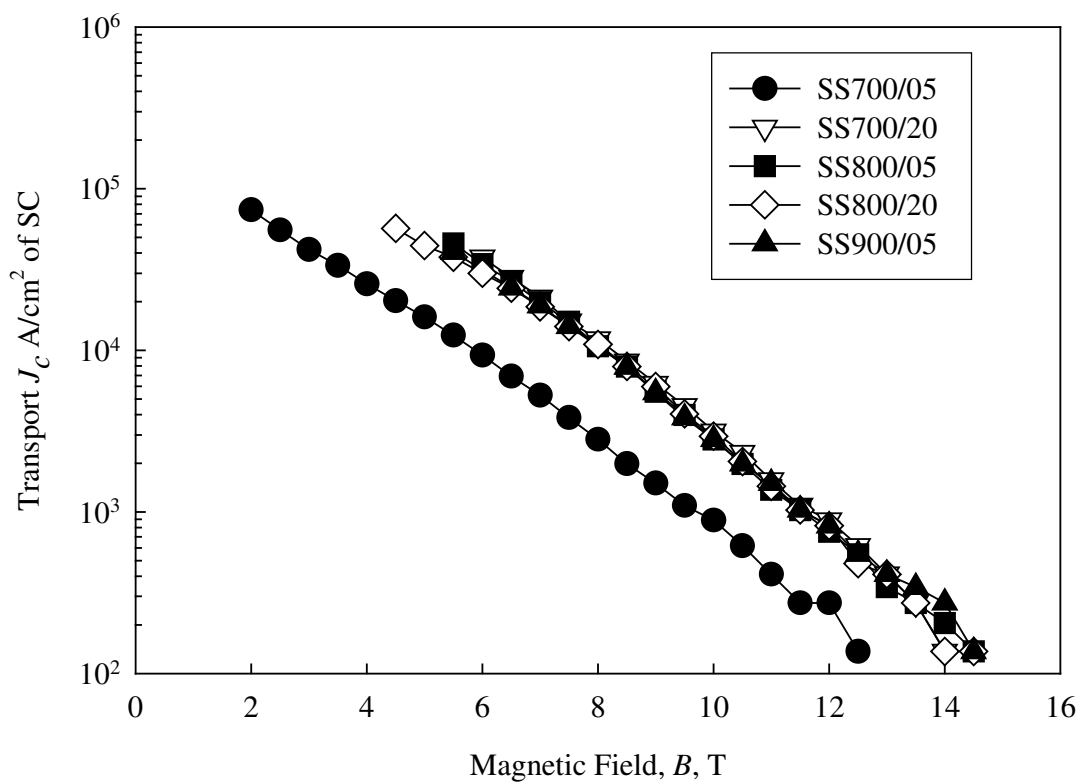

**Figure 3. SUMPTION**

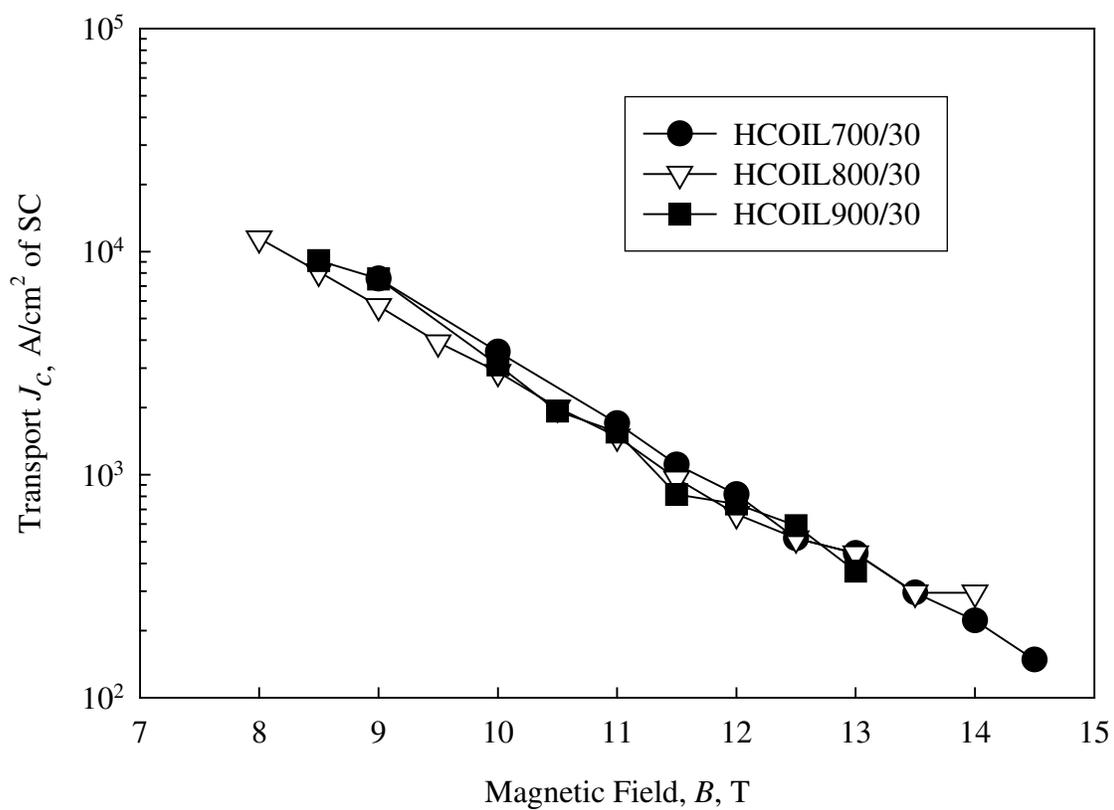

**Figure 4. SUMPTION**